\documentclass{article}

\usepackage[psamsfonts]{amssymb}
\usepackage{graphicx}
\usepackage{bm}
\usepackage{amsmath}

\def\be{\begin{equation}} \def\ee{\end{equation}}
\def\bea{\begin{eqnarray}} \def\eea{\end{eqnarray}}

\begin{document}

\pagestyle{plain}

\def\e{{\rm e}} \def\ie{{\it i.e.}, }
\def\d{\mathrm{d}} \def\id{\mathbf{1}}
\def\ir{\mathrm{i}\;\!}

~~
\vspace{2cm}

\begin{center}

{\Large {\bf Particle Dynamics And Emergent Gravity}}

\vspace{1.5cm}

Amir H. Fatollahi

\vspace{0.5cm}

{\it Department of Physics, Alzahra University, Tehran 19938-91167, Iran}

\vspace{.5cm}

{\sl ahfatol@gmail.com}

\vskip 1.5 cm
\end{center}

\begin{abstract}
The emergent gravity proposal is examined within the
framework of noncommutative QED/gravity correspondence
from particle dynamics point of view.
\end{abstract}

\vspace{1.0cm}

\newpage
\section{Introduction}
There have been arguments supporting the idea that the quantum theory of gravity,
as the theory governing the quantum fluctuations of spacetime, might be formulated most
naturally on noncommutative~(NC) spacetime \cite{doplic,madore1}.
One of the proposals in this context is that, maybe gravity should
be considered as an emergent phenomena rather than a fundamental one, the so-called
``emergent gravity" scenario \cite{rivel,muthu,yang,stein}.
Accordingly, it might be likely that the gravity effects are simply an interpretation of
the interaction of NC gauge fields and matter.
In particular, it is observed that, at least in some available expansion in NC parameter,
a proper rearrangement of interaction terms of a NC
U(1) background with matter field can be interpreted as a free theory but in curved background
\cite{rivel,muthu,yang,stein}.
In \cite{stein} the idea is pushed forward from the classical level to quantum one, leading to
the novel observation that the so-called UV/IR mixing phenomena \cite{uvir} plays an essential
role once one wants to make correspondence between the one-loop effective action in gauge theory
and gravity sides. In another observation of this kind it is seen that the relative dynamics of two
massive NC photons at low energy is described by a free theory but in a modified metric \cite{bound}.
This sounds it is highly expected that the gravity effects for both the ordinary matter as well as
the massless NC photons, which eventually come as the dynamical degrees of freedom of the resulting
theory of gravity, are governed by the same NC~U(1) theory.

In this note the aim is to present working examples by which the basic features and the extent
of the proposed NC~U(1)/gravity correspondence can be understood from the particle dynamics point of view.
In particular it is shown that how the NC~U(1) backgrounds that satisfy certain condition
can turn to a curved background, though in some special gauges of diffeomorphism transformations.
The NC~U(1)/gravity correspondence is examined by the classical equation of motion of
the particle in the final section.

Here we consider the canonical NC spacetime, whose coordinates satisfy the algebra
\be\label{1}
\big[\hat{x}^\mu,\hat{x}^\nu\big]=\ir\theta^{\mu \nu}\,\id,
\ee
in which $\theta^{\mu \nu}$ is an antisymmetric constant tensor and $\id$
represents the unit operator. It has been understood that
the longitudinal directions of D-branes in the presence
of a constant B-field background appear to be noncommutative,
as seen by the ends of open strings \cite{nc-can}.
It is understood that theories on canonical NC spacetime are defined
by actions that are essentially the same as in ordinary spacetime, with the
exception that the products between fields are replaced by $\star$-products,
defined for two functions $f$ and $g$ by
\bea
(f\star g)(x)=\exp\Big(\frac{\ir\theta^{\mu\nu}}{2}\,\partial_{x_\mu}\partial_{y_\nu}\Big)f(x)\, g(y)\Big|_{y=x}.
\eea

\section{U(1) Background As Curved Background}
The starting point is the action
\bea\label{2}
S=\int \d^D x \, \d t \Bigg(
\displaystyle{\frac{1}{2\,m}}
\Big(\hbar {\bm{\nabla}} \psi-\ir \big[\mathbf{A},\psi\big]_\star\Big) \cdot
\Big(\hbar \bm{\nabla} \psi^*-\ir \big[\mathbf{A},\psi^*\big]_\star\Big)
-\ir \psi^* \Big( \hbar\,\partial_t \psi -\ir \big[A_0,\psi\big]_\star\Big)
\Bigg)
\eea
in which $\big[a\, ,b\big]_\star:= a \,\star\, b - b\, \star\, a$, and
$\psi^*$ is the complex conjugate of $\psi$. The equation of motion by the
above action presents the quantum mechanics of a particle in presence of the
NC~U(1) background $(A_0,\mathbf{A})$ in the Schr\"{o}dinger picture.
We mention that here the matter $\psi$ interacts
in the sense of adjoint representation with the background. This kind of interaction
is absent in U(1) theory on ordinary spacetime, as there the product is commutative,
opposed to $\star$-product here. In the following we set $\hbar=1$ for sake of simplicity; whenever
needed $\hbar$'s can be restored by dimensional considerations.

The action above is invariant under the gauge transformations:
\bea\label{3}
\psi  &\to& \psi' = U \star \psi \star U^* \cr
\mathbf{A} &\to& \mathbf{A}' = U \star \mathbf{A} \star U^* + \ir\, U \star \bm{\nabla} U^* \cr
A_0 &\to& A_0' = U\star A_0 \star U^* + \ir\, U \star \partial_t U^*
\eea
in which $U$ is a $\star$-phase defined by
\bea
U=\exp_\star(\,\ir \Lambda) = 1+ \ir \Lambda -\frac{1}{2} \Lambda \star \Lambda +\cdots .
\eea
with $\Lambda$ as an arbitrary function. One easily can show that $U\star U^* = U^* \star U =1$.

Here we consider the scattering of the particle by the given background at the lowest level.
In the following we assume that 1)~$A_0=0$,
and 2)~the noncommutativity is restricted to spatial directions,
$\theta^{0i}=0$. We take that the incoming and outgoing particles have momenta
$\mathbf{p}_1$ and $\mathbf{p}_2$, respectively.
The expression for the transition amplitude can simply be obtained via the Feynman rules of
the field theory set up of the problem
\bea\label{4}
T_{1\to\, 2}= \frac{\ir}{m}  \sin\!\Big(\frac{\mathbf{p}_1 \ltimes \mathbf{p}_2}{2}\Big)~
(\mathbf{p}_1+\mathbf{p}_2)\cdot \widetilde{\mathbf{A}}(\mathbf{q},\omega)
\eea
in which $\mathbf{k} \ltimes \mathbf{l}=\theta^{ij} k_i\,l_j$, and
$\widetilde{\mathbf{A}}(\mathbf{q},\omega)$ is the Fourier transform of the background $\mathbf{A}(\mathbf{x},t)$,
\bea\label{5}
\widetilde{\mathbf{A}}(\mathbf{q},\omega)=\int \d^D x ~ \d t
~\e^{\ir \mathbf{q}\cdot \mathbf{x} -\ir \omega t}\, \mathbf{A}(\mathbf{x},t)
\eea
with $\mathbf{q}=\mathbf{p}_2-\mathbf{p}_1$ and $\omega=E_2-E_1$, as transferred momenta and energy, respectively.
We mention a certain combination of the incoming and outgoing particles's momenta determine the
strength of the interaction. In fact in higher orders higher powers of momenta come in the expression. So,
as we are working at the lowest order, we take the sine equal to its argument, getting
\bea\label{6}
T_{1\to\, 2}= \frac{\,\ir}{2\,m} \, (\mathbf{p}_1 \ltimes \mathbf{p}_2)~
(\mathbf{p}_1+\mathbf{p}_2)\cdot \widetilde{\mathbf{A}}(\mathbf{q},\omega).
\eea
Now in quantum mechanical interpretation of the problem on ordinary spacetime,
the result above corresponds to the first order Born approximation expression
\bea\label{7}
T_{1\to\, 2}=  \int \d t ~\e^{-\ir (E_2-E_1) t}\;\big\langle \, \mathbf{p}_1 \big|
\hat{V}(\hat{\mathbf{x}},\hat{\mathbf{p}},t) \big| \, \mathbf{p}_2 \big\rangle
\eea
with
\bea\label{8}
\hat{H} = \frac{\hat{\mathbf{p}}\cdot \hat{\mathbf{p}}}{2\,m} + \hat{V}(\hat{\mathbf{x}},\hat{\mathbf{p}},t),~~~~~~
\big\langle \, \mathbf{x}\, \big| \, \mathbf{p} \big\rangle = \e^{\ir \mathbf{p}\cdot \mathbf{x}}\cr\cr
\big[\hat{x}^i, \hat{x}^j \big]=\big[\hat{p}_i, \hat{p}_j \big]=0, ~~~~~~~
\big[\hat{x}^i, \hat{p}_j \big]= \ir \hbar \,\delta^i_j\, \id.
\eea
By comparing (\ref{6}) and (\ref{7}), recalling the Hermiticity of Hamiltonian,
and with $\hat{\mathbf{A}}=\hat{\mathbf{A}}(\hat{\mathbf{x}},t)$, one has the following
expression for $\hat{V}$
\bea\label{9}
\hat{V}(\hat{\mathbf{x}},\hat{\mathbf{p}},t) = \frac{\,\ir}{4\,m}\, \theta^{ij}
\bigg\{ \hat{p}_i \big(\hat{\mathbf{A}}\cdot\hat{\mathbf{p}}+\hat{\mathbf{p}}\cdot\hat{\mathbf{A}}\big) \, \hat{p}_j
- \hat{p}_j \big(\hat{\mathbf{A}}\cdot\hat{\mathbf{p}}+\hat{\mathbf{p}}\cdot\hat{\mathbf{A}}\big) \, \hat{p}_i \bigg\}
\eea
Using the identities $\hat{p}_i \hat{A}_j = \hat{A}_j \hat{p}_i - \ir \partial_i \hat{A}_j$,
and $\theta^{ij} \,\hat{p}_i\, \hat{p}_j = 0$, one finds
\bea\label{10}
\hat{V}(\hat{\mathbf{x}},\hat{\mathbf{p}},t) = \frac{1}{2\,m}\,
\big(\hat{\gamma}^{ij}\, \hat{p}_i \, \hat{p}_j -  \ir \,\partial_i \hat{\gamma}^{ij}\, \hat{p}_j \big)
\eea
in which
\bea\label{11}
\hat{\gamma}^{ij}(\hat{\mathbf{x}},t) = \theta^{kj} \partial_k \hat{A}^i + \theta^{ki} \partial_k \hat{A}^j
=\hat{\gamma}^{j\,i}(\hat{\mathbf{x}},t)
\eea
satisfying
\bea\label{11.1}
\partial_i\partial_j \hat{\gamma}^{ij} = 0
\eea
using $\theta^{nl} \partial_n \partial_l=0$. The Hamiltonian then takes the form of
\bea\label{12}
\hat{H} = \frac{1}{2\,m}
\big( \hat{g}^{ij} \, \hat{p}_i \, \hat{p}_j -  \ir \,\partial_i \hat{g}^{ij}\, \hat{p}_j \big)
\eea
with $\hat{g}^{ij}(\hat{\mathbf{x}},t)=\delta^{ij} + \hat{\gamma}^{ij}(\hat{\mathbf{x}},t)$.
We mention that the above Hamiltonian, despite presence of combinations of position and momentum
operators, is free from ordering ambiguity. The above Hamiltonian in the
position basis takes the form of
\bea\label{13}
H_{\rm {\bf x}\,basis} = \frac{-1}{2\,m}
\big( g^{ij} \, \partial_i \, \partial_j + \partial_i g^{ij}\, \partial_j \big)
\eea
According to recipe the Hamiltonian operator of a free particle in position basis
is given by the Laplacian constructed by the metric $g^{ij}(\mathbf{x},t)$,
\bea\label{14}
H_{\rm {\bf x}\,basis}  =  - \frac{1}{2\, m} \, \nabla^2 =
 - \frac{1}{2\, m} \,\frac{1}{\sqrt{g}}\, \partial_i \big(\sqrt{g}\, g^{ij}\, \partial_j \big)
\eea
in which $g = \det g_{ij}$. So interpreting (\ref{13}) as the Hamiltonian of kind (\ref{14})
gives the uni-modular condition $\det g_{ij}=1$, which for acceptable order of $\theta$ says
\bea\label{15}
{\rm Tr}\, \gamma^{ij} = \gamma^i_{~i}~ \propto ~\theta^{ij} F_{ij} = 0
\eea
with $F_{ij}= \partial_i A_j - \partial_j A_i$. We mention that the condition above can not be
considered as a gauge fixing one, because $F_{ij}$ is gauge invariant, at least at this order of $\theta$.
So according to the construction above, for NC~U(1) background $\mathbf{A}(\mathbf{x},t)$
that satisfies (\ref{15}) the dynamics is described by the motion of particle in presence of the effective metric
\bea\label{18}
g^{ij}  = \delta^{ij} + \theta^{kj} \partial_k A^i + \theta^{ki} \partial_k A^j,
\eea
satisfying the conditions
\bea\label{19}
\partial_i\partial_j g^{ij} = 0, ~~~~~~~~~~\det g_{ij} = 1.
\eea

It is useful to compare the construction above with that of \cite{rivel}, in which the case of
a massless scalar field $\widehat{\varphi}$ is considered in presence of the NC background
$\widehat{A}_\mu$. Since the action has only one term, one can use the
the first order Seiberg-Witten map
\bea\label{16}
\widehat{A}_\mu &=& A_\mu - \frac{1}{2} \theta^{\alpha\beta} A_\alpha (\partial_\beta A_\mu + F_{\beta\mu}) \cr \cr
\widehat{\varphi} &=& \varphi - \theta^{\alpha\beta} A_\alpha \partial_\beta \varphi
\eea
that turns the action to the one for the scalar field $\varphi$ in curved background given by
the $A_\mu$-dependent effective metric $G^{\mu\nu}$ \cite{rivel}, for which we have
\bea\label{17}
\det G_{\mu\nu} - \det \eta_{\mu\nu} ~\propto ~ \frac{D-3}{2}~ \theta^{\alpha\beta} F_{\alpha\beta} + O(\theta^2),
\eea
in $D+1$ space-time dimensions. We see that in 3+1 dimensions, the uni-modular condition is satisfied
automatically.


\section{Classical Dynamics In NC U(1) Background}
In this section we derive the classical equation of motion of a particle with NC charge.
The NC~QED is given by the action ($\hbar=c=1$)
\bea
S=\int \d^D x\,\d t\; \bigg(\!\!-\frac{1}{4}\,F_{\mu\nu}F^{\mu\nu}
+\ir\, \overline{\psi}\,\gamma^{\mu}\Big(\partial_\mu \psi - \ir \big[ A_{\mu}, \psi\big]_\star\Big)
- m \, \overline\psi\,\psi\bigg), \\ \cr
\eta^{\mu\nu}={\rm diag}(+1,-1,\cdots,-1),~~\gamma^\mu\gamma^\nu + \gamma^\nu\gamma^\mu = 2\,\eta^{\mu\nu}, \nonumber
\eea
in which $\overline \psi = \psi^\dagger \gamma^0$, and field strength is defined by
\bea
F_{\mu\nu}=\partial_{\mu}A_{\nu}-\partial_{\nu}A_{\mu}-\ir \big[A_{\mu},A_{\nu}\big]_\star.
\eea
The action above is invariant under the gauge transformation (\ref{3}).
Under the gauge transformation, the field strength transforms as
\bea
F_{\mu\nu}\to F^{\prime}_{\mu\nu}=U\star F_{\mu\nu}\star U^*
\eea
We mention that the transformations of gauge fields as well as the field strength
of NC~U(1) theory, together with self-interaction terms in the pure gauge sector in
action above, shows that it should be regarded as a non-Abelian gauge theory.
Based on these facts, here we use the Wong's approach, originally adopted
for particles with non-Abelian charge \cite{wong}, to derive the classical equations
of motion of charges in presence of NC~U(1) background. This formalism is presented
for charges in fundamental representation in \cite{fat-moh}. Here we consider charges
in adjoint representation, and give a presentation appropriate for emergent gravity
interpretation of the result. At the first order of NC parameter the Lagrangian takes the form of
\bea
L=-\frac{1}{4}\,F^{\mu\nu}F_{\mu\nu}+\ir\,\overline {\psi}\gamma^{\mu}\partial_{\mu}\psi
+\ir\,\overline {\psi}\gamma^{\mu}\theta^{\alpha\beta}\partial_{\alpha}A_{\mu}\partial_{\beta}\psi-m\overline {\psi}\psi + O(\theta^2),
\eea
by which the equation of motion for $\psi$ is obtained to be:
\bea
\ir\,\gamma^{0}\partial_{0}\psi+\ir\,\gamma^{i}\partial_{i}\psi+\ir\,\gamma^{\mu}\theta^{\alpha\beta}
\partial_{\alpha}A_{\mu}\partial_{\beta}\psi-m\psi=0
\eea
As previous section we assume 1) noncommutativity is just in spatial directions: $\theta^{0i}=0$, 2) $A_0=0$.
So, the above equation appears in the form:
\bea
\ir\,\gamma^{0}\partial_{0}\psi+\ir\,\gamma^{i}\partial_{i}\psi
+\ir\,\theta^{ij}\gamma^{k}\partial_{i}A_{k}\partial_{j}\psi -m\psi=0
\eea
Taking above as a Schr\"{o}dinger equation we read the corresponding Hamiltonian as
\bea
\hat{H}=\alpha^{i}\hat{p}_i+\theta^{ij}\alpha^k\,\partial_iA_k\,\hat{p}_j+m\gamma^0
\eea
in which $\alpha^{k}=\gamma^0 \gamma^k$.
The Heisenberg equations of motion are derived for the operators as well:
\bea
\dot{\hat{x}}^{k}\!\!\!&=&\!\!\!\ir\,[\hat{H},\hat{x}^{k}]=\alpha^{k}+\theta^{ik}\alpha^{j}\,\partial_{i}A_{j}\\
\dot{\hat{p}}_{k}\!\!\!&=&\!\!\!\ir\,[\hat{H},\hat{p}_{k}]=-\,\theta^{ij}\alpha^{l}\,\partial_{k}\partial_{i}A_{l}\,\hat{p}_j
\eea
From the first equation we have
$\alpha^k=\dot{\hat{x}}^{k}-\theta^{ik}\,\dot{\hat{x}}^{j}\,\partial_{i}A_{j}+O(\theta^2)$,
which gives by second equation:
\bea
\dot{\hat{p}}_{k}=-\,\theta^{ij}\,\dot{\hat{x}}^{l}\,\partial_{k}\partial_{i}A_{l}\,\hat{p}_j  +O(\theta^2).
\eea
One might take above as the equation of motion $p_k$ in the curved background as
\bea
p_k = m \, \tilde{g}_{kl}\, \dot{x}^l,~~~~~~~ \tilde{g}_{nl}(\mathbf{x},t)= \delta_{nl} - \tilde{\gamma}_{nl}(\mathbf{x},t)
\eea
in which we have gone to the classical regime by dropping the hats. By this interpretation one finds
\bea
\ddot{x}^{\,l} + \frac{1}{2}\, \delta^{lk}\,
\Big(\partial_i \tilde{\gamma}_{jk} + \partial_j \tilde{\gamma}_{ik}
- \partial_{k} \tilde{\gamma}_{ij}   \Big) \, \dot{x}^i \dot{x}^j + \partial_0 \tilde{\gamma}^{il}\dot{x}_i
+ O(\tilde{\gamma}^2)= 0
\eea
with $\tilde{\gamma}^{ij}=\theta^{kj} \partial_k A^i + \theta^{ki} \partial_k A^j$,
as the same coming in metric (\ref{18}). We mention that above result is simply
the dynamics of a particle in presence of a time-dependent metric.

\vspace{.5cm}
{\bf Acknowledgement:} This work is partially supported by the Research Council of Alzahra University.

\end{document}